\documentclass[fleqn,10pt]{article}
\usepackage{amssymb}
\usepackage{amsmath}
\usepackage{amsfonts}
\usepackage{graphicx}

\renewcommand{\theequation}{\arabic{section}.\arabic{equation}}
\newcommand{\R}{\,{\rm{Re}}\,}

\begin{document}
\title{\textbf{Soliton Equations with Self-Consistent Sources}}

\author{ {\bf  Ratbay Myrzakulov\footnote{Email: rmyrzakulov@gmail.com}} \\
    {\small {\it
    Department of General and Theoretical Physics, Eurasian National University,
    Astana 010008, Kazakhstan}} }

\date{}
\maketitle
\renewcommand{\theequation}{\arabic{section}.\arabic{equation}}

\begin{abstract}
We consider some soliton equations with  self-consistent
sources. A brief review of main SESCS is presented. In particular we construct the Heisenberg ferromagneic equation with  self-consistent
sources (HFESCS) which is integrable. The corresponding Lax representation is presented.  Some  properties of HFESCS 
are analyzed. The relation between soliton equations with  self-consistent potentials and soliton equations with  self-consistent
sources is studied.

\end{abstract}

\hskip\parindent

\tableofcontents

\section{Introduction}
\setcounter{equation}{0} \hskip\parindent

 Soliton equations with self-consistent
sources (SESCS) were  proposed in \cite{mel89a}. They  describe nonlinear waves and have important applications in  plasma
physics,  hydrodynamics,  solid state physics etc. As example, we can mention that one of important SESCS - the nonlinear  Schrodinger
equation with self-consistent sources (NLSESCS),   describes the nonlinear interaction of an ion acoustic wave in a two component homogeneous plasma  with the electrostatic highfrequency
wave \cite{Claude91}. Other important SESCS, the KP
equation with self-consistent sources (KPESCS) represents the interaction of a short wave
packet propagating on the x-y plane with the long wave   at some angle to each other (see \cite{Mel'nikov87} and the references therein). As the third example, we can mention that the interaction of long and short capillary-gravity waves describes by the   the KdV equation with self-consistent sources (KdVESCS) \cite{Leon90(1)}. In such models,  the sources may change some properties of the physical system, for example, 
the velocities of the solitons. In literature, some explicit
solutions of SESCS  like: solitons, positons and  negatons  were obtained.  We note that for the given a soliton equation, its version with self-consistent sources is not unique. Some SESCS and their properties were studied in the literature (see e.g. refs. \cite{Lin1}-\cite{Lin3} and references therein).

In this paper, we present the main SESCS with their Lax representations (LR) in continuous, ddiscrete and dispersionless cases. Then we give the Heisenberg ferromagneic equation with  self-consistent
sources (HFESCS). This HFESCSC   is integrable.

The structure of the paper is as follows. In Sec.2, we present a brief review of the main SESCS for three cases - continuous, discrete and dispersionless. Then in Sec.3, we construct the integrable HFESCS and its some properties are analysysed. In Sec. 4, we present some integrable generalizations of HFESCS. In Sec. 5, we study the relation between soliton equations with  self-consistent potentials and soliton equations with  self-consistent
sources. We summarize the main results of this paper in Sec. 5. 

\section{Brief review of soliton equations with self-consistent sources}

To make the paper self-contained, we first briefly recall the main soliton equations with self-consistent sources. Then we will give the HFESCS,  its Lax reresentation (LR), gauge equivalent counterpart and some generalizations. 

\subsection{Continuous (1+1)-dimensional equations}

\subsubsection{KdV equation with self-consistent sources}
The KdVESCS looks like (see e.g. \cite{0304030})
\begin{subequations}
\label{2.1}
\begin{equation}
\label{2.1a}
  u_t+6uu_x+u_{xxx}+4\sum_{j=1}^n\varphi_j\varphi_{j,x}=0,
\end{equation}
\begin{equation}
\label{2.1b}
  \varphi_{j,xx}+(\lambda_j+u)\varphi_j=0,\quad j=1,\ldots,n,
\end{equation}
\end{subequations}
where $\lambda_j$ = real constants.  The corresponding LR is given by
\begin{subequations}
\label{2.2}
\begin{equation}
\label{2.2a}
  \phi_{xx}+(\lambda+u)\phi=0,
\end{equation}
\begin{equation}
\label{2.2b}
  \phi_t=A_n(\lambda,u,\Phi_n)\phi,
\end{equation}
\end{subequations}
where 
$$A_n(\lambda,u,\Phi_n)\phi=u_x\phi+(4\lambda-2u)\phi_x
+\sum_{j=1}^n\frac{\varphi_j}{\lambda_j-\lambda}W(\varphi_j,\phi).$$
Here  $W(\varphi_j,\phi)\equiv\varphi_j\phi_x-\varphi_{j,x}\phi$ is
the usual Wronskian determinant and $\Phi_n=(\varphi_1,\ldots,\varphi_n)$.
\subsubsection{mKdV equation  with self-consistent sources}

The mKdV equation with
self-consistent sources (mKdVSCS)  reads as \cite{0806.2251}
\begin{subequations}
\label{2.3}
  \begin{align}\label{2.3a}
   u_{t}=\frac{u_{xxx}}{4}-\frac{3}{2}u^{2}u_{x}+\sum\limits_{j=1}^{N}(\phi_{1j}\phi_{2j})_{x},
	\end{align}
  \begin{align}\label{2.3b}
	\phi_{1jx}=\lambda_{j}\phi_{2j}
-u\phi_{1j},~\phi_{2jx}=\lambda_{j}\phi_{1j} +u
\phi_{2j},~~~~j=1,\cdots, N.
  \end{align}
\end{subequations}
 
\subsubsection{NLS equation  with self-consistent sources}

The nonlinear Schrodinger  equation  with self-consistent sources (NLSESCS) is  given  by \cite{0412071}
\begin{subequations}
\label{2.4}
\begin{equation}
\label{2.4a}
\varphi_{j,x}=U(\lambda_j,q,r)\varphi_j,\quad \varphi'_{j,x}=U(\lambda'_j,q,r)\varphi'_j,
\quad j=1,\ldots,m,
\end{equation}
\begin{equation}
\label{2.4b}
\phi_{j,x}=U(\zeta_j,q,r)\phi_j,\quad j=1,\ldots,n,
\end{equation}
\begin{equation}
\label{2.4c}
q_t=-i(q_{xx}-2q^2r)+\sum_{j=1}^{m}\left[(\varphi_j^{(1)})^2+({\varphi_j'}^{(1)})^2\right]+\sum_{j=1}^n(\phi_j^{(1)})^2,
\end{equation}
\begin{equation}
\label{2.4d}
r_t=i(r_{xx}-2qr^2)+\sum_{j=1}^{m}\left[(\varphi_j^{(2)})^2+({\varphi_j'}^{(2)})^2\right]+\sum_{j=1}^n(\phi_j^{(2)})^2.
\end{equation}
\end{subequations}
The corresponding LR reads as
\begin{equation}
\label{2.5}
\psi_x=U(\lambda,q,r)\psi,\quad \psi_t=V(\lambda,q,r)\psi+\sum_{j=1}^m
\left[\frac{H(\varphi_j)}{\lambda-\lambda_j}+\frac{H(\varphi'_j)}{\lambda-\lambda'_j}\right]\psi
+\sum_{j=1}^n\frac{H(\phi_j)}{\lambda-\zeta_j}\psi.
\end{equation}

Let us also now we present two reductions of the NLSESCS. 

i) Let
\begin{subequations}
\label{2.6}
\begin{equation}
\label{2.6a}
r=q^*,\quad \lambda'_j=-\lambda_j^*\quad \varphi'_j=\pm S_+\varphi_j,\quad j=1,\ldots,m,
\end{equation}
\begin{equation}
\label{2.6b} \ \R\zeta_j=0,\quad
{\phi_j^{(2)}}^*=\phi_j^{(1)}\equiv w_j,\quad j=1,\ldots,n.
\end{equation}
\end{subequations}
Then we obtain the following equations
\begin{subequations}
\label{2.7}
\begin{equation}
\label{2.7a}
\varphi_{j,x}=U(\lambda_j,q,q^*)\varphi_j,\quad j=1,\ldots,m,
\end{equation}
\begin{equation}
\label{2.7b} w_{j,x}=\zeta_jw_j+qw^*_j,\ (\R\zeta_j=0),\quad
j=1,\ldots,n,
\end{equation}
\begin{equation}
\label{2.7c}
q_t=i(2|q|^2q-q_{xx})+\sum_{j=1}^m\left[(\varphi_j^{(1)})^2+({\varphi_j^{(2)}}^*)^2\right]
+\sum_{j=1}^n w_k^2
\end{equation}
\end{subequations}
with LR
\begin{equation}
\label{2.8} \psi_x=U(\lambda,q,q^*)\psi,\quad
\psi_t=V(\lambda,q,q^*)\psi+\sum_{j=1}^m
\left[\frac{H(\varphi_j)}{\lambda-\lambda_j}+\frac{H(S_+\varphi_j)}{\lambda+\lambda_j^*}\right]\psi
+\sum_{j=1}^n\frac{H((w_j,w_j^*)^T)}{\lambda-\zeta_j}\psi.
\end{equation}

ii) Now we take the reduction
\begin{equation}
\label{2.9}
r=-q^*,\quad \lambda'_j=-\lambda_j^*\quad \varphi'_j=\pm iS_-\varphi_j,\quad j=1,\ldots,m.
\end{equation}
Then we come to the system
\begin{subequations}
\label{2.10}
\begin{equation}
\label{2.10a}
\varphi_{j,x}=U(\lambda_j,q,-q^*)\varphi_j,\quad j=1,\ldots,m,
\end{equation}
\begin{equation}
\label{2.10b}
q_t=i(-2|q|^2q-q_{xx})+\sum_{j=1}^m\left[(\varphi_j^{(1)})^2-(\varphi_j^{(2)*})^2\right]
\end{equation}
\end{subequations}
with the LR of the form
\begin{equation}
\label{2.11}
\psi_x=U(\lambda,q,-q^*)\psi,\quad \psi_t=V(\lambda,q,-q^*)\psi+\sum_{j=1}^m
\left[\frac{H(\varphi_j)}{\lambda-\lambda_j}
-\frac{H(S_-\varphi_j)}{\lambda+\lambda_j^*}\right]\psi.
\end{equation}

\subsubsection{Camassa-Holm equation  with self-consistent sources}
The Camassa-Holm equation  with self-consistent sources (CHESCS) is defined as follows \cite{0811.2552}
\begin{subequations}
\label{2.12}
\begin{eqnarray}\label{2.12a}
q_t&=&-J(\frac{\delta H_0}{\delta
q}-2\sum_{j=1}^N\frac{\delta\lambda_j}{\delta q})\notag\\
&=&-(q\partial+\partial
q)(u+2\sum_{j=1}^N\lambda_j\varphi_j^2)\notag\\
&=&-2qu_x-uq_x+\sum_{j=1}^N(-8\lambda_jq\varphi_j\varphi_{jx}-2
\lambda_jq_x\varphi_j^2),
\end{eqnarray}
\begin{eqnarray}\label{2.12b}
\varphi_{j,xx}&=&(\lambda_j q+\frac{1}{4})\varphi_j,\quad
j=1,\cdots,N.
\end{eqnarray}
\end{subequations}
Equivalently this system can be  written as
\begin{subequations}
\label{2.13}
\begin{eqnarray}\label{2.13a}
q_t&=&-2qu_x-uq_x+\sum_{j=1}^N[(\varphi_j^2)_x-(\varphi_j^2)_{xxx}],
\end{eqnarray}
\begin{eqnarray}\label{2.13b}
\varphi_{j,xx}&=&(\lambda_j q+\frac{1}{4})\varphi_j,\quad
j=1,\cdots,N.
\end{eqnarray}
\end{subequations}

The LR of the CHESCS has the form
\begin{subequations}
\label{2.14}
 \begin{align}\label{2.14a}
\varphi_{xx}=(\lambda q+\frac{1}{4})\varphi,
 \end{align}
\begin{align}\label{2.14b}
\varphi_{t}=-\frac{1}{2}B_{x}\varphi+B\varphi_{x},
 \end{align}
\begin{align}\label{2.14c}
B=\frac{1}{2\lambda}-u+\sum\limits_{j=1}^{N}\frac{\alpha_{j}f(\varphi_{j})}{\lambda-\lambda_{j}}+
\sum\limits_{j=1}^{N}\beta_{j}f(\varphi_{j}),
 \end{align}
\end{subequations}
where  $f(\varphi_{j})$ is some undetermined function of $\varphi_{j}$.
From these equations we get
\begin{equation}
\label{2.15}
\lambda q_{t}=LB+\lambda(2B_{x}q+Bq_{x}),
\end{equation}
where $L=-\frac{1}{2}\partial^{3}+\frac{1}{2}\partial$. So finally we obtain
$$\lambda q_{t}=-\frac{1}{2}\sum\limits_{j=1}^{N}\frac{\alpha_{j}}
{\lambda-\lambda_{j}}[f^{'''}\varphi_{jx}^{3}+3(f^{''}\varphi_{j}-f')(\lambda_{j}q+
\frac{1}{4})\varphi_{jx}+\lambda_{j}q_{x}(f'\varphi_{j}-2f)]$$
$$+[-2qu_{x}-uq_{x}+\sum
\limits_{j=1}^{N}\beta_{j}(2q\varphi_{jx}f'+q_{x}f)]\lambda-\frac{1}{2}
\sum\limits_{j=1}^{N}\beta_{j}[f^{'''}\varphi_{jx}^{2}+(3f^{''}\varphi_{j}+f')$$
\begin{equation}
\label{2.16} \times (\lambda_{j}q+
\frac{1}{4})\varphi_{jx}+\lambda_{j}f'q_{x}\varphi_{j}-f'\varphi_{j}]+\sum
\limits_{j=1}^{N}\alpha_{j}(q_{x}f+2qf'\varphi_{jx}).
\end{equation}
\subsubsection{Degasperis-Procesi equation  with self-consistent sources}
The Degasperis-Procesi equation  with self-consistent sources (DPESCS) has the form \cite{0807.0085}
\begin{subequations}
\label{2.17}
\begin{eqnarray}\label{2.17a}
m_t&=&-um_x-3u_xm-\frac{1}{6}\sum_{j=1}^n\partial(1-\partial^2)(4-\partial^2)(\lambda_jq_jr_j),
\end{eqnarray}
\begin{eqnarray}\label{2.17b}
q_{j,xxx}&=&q_{j,x}-m\lambda_jq_j,
\end{eqnarray}
\begin{eqnarray}\label{2.17c}
r_{j,xxx}&=&r_{j,x}+m\lambda_jr_j,\quad j=1,\cdots, n.
\end{eqnarray}
\end{subequations}
The LR  of the DPESCS reads as
\begin{subequations}
\label{2.18}
\begin{eqnarray}\label{2.18a}
\psi_{xxx}&=&\psi_x-m\lambda\psi,
\end{eqnarray}
\begin{eqnarray}\label{2.18b}
\psi_t&=&-\frac{1}{\lambda}\psi_{xx}-u\psi_x+(u_x+\frac{2}{3\lambda})\psi\notag\\
&&+(\sum_{j=1}^n{\frac{1}{6}\frac{\lambda\lambda_j^2}{\lambda_j^2-\lambda^2}(3\lambda(q_jr_{j,xx}-q_{j,xx}r_j)-4\lambda_jq_jr_j-2\lambda_j(q_jr_j)_{xx})})\psi\notag\\
&&+(\sum_{j=1}^n{-\frac{1}{2}\frac{\lambda\lambda_j^2}{\lambda_j^2-\lambda^2}(\lambda(q_jr_{j,x}-q_{j,x}r_j)+\lambda_j(q_jr_j)_x)})\psi_x\notag\\
&&+(\sum_{j=1}^n{\frac{\lambda\lambda_j^3}{\lambda_j^2-\lambda^2}q_jr_j})\psi_{xx}.
\end{eqnarray}
\end{subequations}

\subsubsection{AKNS equation with self-consistent sources}
The AKNS equation with self-consistent sources (AKNSESCS) is given by \cite{0412071}
\begin{equation}
\label{2.19a}
q_{t}=-i(q_{xx}-2q^2r)+\sum_{j=1}^{n}(\varphi_j^{(1)})^2,
\end{equation}
\begin{equation}
r_{t}=i(r_{xx}-2qr^2)+\sum_{j=1}^{n}(\varphi_j^{(2)})^2,
\end{equation}
\begin{equation}
\label{2.19b}
\varphi_{j,x}
=\left(\begin{array}{cc}
-\lambda_j&q\\
r&\lambda_j
\end{array}\right)
\varphi_j,\quad j=1,\cdots,n,
\end{equation}
where $\varphi_j=(\varphi_j^{(1)},\varphi_j^{(2)})^T$. The LR for the AKNSSCS  looks like 
\begin{equation}
\label{2.20a}
\psi_x=U\psi,
\end{equation}
\begin{equation}
\label{2.20b}
\psi_{t_s}=R^{(n)}\psi,
\end{equation}
where
\begin{equation}
\label{2.20b}
R^{(n)}=V+\sum_{j=1}^n\frac{H(\varphi_j)}{\lambda-\lambda_j}
\end{equation}
and 
\begin{equation}
U=\left(\begin{array}{cc}
-\lambda&q\\
r&\lambda
\end{array}\right),\quad V=i\left(\begin{array}{cc}
-2\lambda^2+qr&2\lambda q-q_x\\
2\lambda r+r_x&2\lambda^2-qr
\end{array}\right),
\quad
H(\varphi_j)=\frac{1}{2}
\left(\begin{array}{cc}
-\varphi^{(1)}_j\varphi^{(2)}_j&(\varphi^{(1)}_j)^2\\
-(\varphi^{(2)}_j)^2&\varphi^{(1)}_j\varphi^{(2)}_j
\end{array}\right).
\end{equation}

\subsection{Continuous (2+1)-dimensional  equations}
\subsubsection{KP equation with self-consistent sources}
The  KP equation with self-consistent
sources (KPESCS) reads as \cite{0510021}
\begin{subequations}
\label{2.21}
\begin{equation}
\label{2.21a}
     [u_{1,t}-3u_1u_{1,x}-\frac{1}{4}u_{1,xxx}+\sum_{i=1}^N(q_ir_i)_x]_x-\frac{3}{4}u_{1,yy} = 0,
\end{equation}
\begin{equation}
\label{2.21b}
    q_{i,y} = q_{i,xx}+2u_1q_i,
\end{equation}
\begin{equation}
\label{2.21c}
    r_{i,y} = -r_{i,xx}-2u_1r_i, \ \ \ i=1,..., N.
\end{equation}
\end{subequations}
Its LR is given by 
\begin{subequations}
\label{2.22}
\begin{equation}\label{2.22a}
\psi_y = \psi_{xx}+2u_1\psi,
\end{equation}
\begin{equation}\label{2.22b}
\psi_t =
\psi_{xxx}+3u_1\psi_x+\frac{3}{2}(u_{1,x}+(\partial^{-1}u_{1,y}))\psi+\sum_{i=1}^Nq_i\partial^{-1}(r_i\psi).
\end{equation}
\end{subequations}
\subsubsection{mKP equation with self-consistent sources}
The mKP equation with self-consistent sources (mKPESCS) reads as \cite{0412066}
\begin{subequations}
\label{2.23}
\begin{equation}
\label{2.23a}
    u_t+u_{xxx}+3\alpha^2D^{-1}(u_{yy})-6\alpha D^{-1}(u_y)u_x-6u^2u_x+4\sum_{i=1}^{N}(\Psi_i\Phi_i)_x=0,
\end{equation}
\begin{equation}\label{2.23b}
    \alpha\Psi_{i,y}=\Psi_{i,xx}-2u\Psi_{i,x},
\end{equation}
\begin{equation}\label{2.23c}
    \alpha\Phi_{i,y}=-\Phi_{i,xx}-2u\Phi_{i,x}.
\end{equation}
\end{subequations}
 Its LR looks like 
\begin{subequations}
\label{2.24}
\begin{equation}
\label{2.24a}
    \alpha\psi_{1,y}=\psi_{1,xx}-2u\psi_{1,x},
\end{equation}
\begin{equation}
\label{2.24b}
    \psi_{1,t}=(A_1(u)\psi_1)+T_N^1(\Psi,\Phi)\psi_1,\ \
    T_N^1(\Psi,\Phi)\psi_1=-4\sum_{i=1}^{N}\Psi_i\int
    \Phi_i\psi_{1,x}{\mathrm{d}}x
\end{equation}
\end{subequations}
and
\begin{subequations}
\label{2.25}
\begin{equation}
\label{2.25a}
    \alpha\psi_{2,y}=-\psi_{2,xx}-2u\psi_{2,x},
\end{equation}
\begin{equation}
\label{2.25b}
    \psi_{2,t}=(A_2(u)\psi_2)+T_N^2(\Psi,\Phi)\psi_2,\ \
    T_N^2(\Psi,\Phi)\psi_2=4\sum_{i=1}^{N}\Phi_i\int
    \Psi_i\psi_{2,x}{\mathrm{d}}x.
\end{equation}
\end{subequations}
\subsubsection{Davey-Stewartson equation with  self-consistent sources}

The  Davey–Stewartson equation with  self-consistent sources (DSESCS)  reads as   \cite{DSSCS}

\begin{equation}\label{2.26}
iu_t+u_{xx}+u_{yy}+u(v_{xx}+v_{yy})=\delta[i(\Psi_{1}\Psi_{2}+\Psi_{1}\Psi_{1})-0.25(\Phi_{1}\Phi_{2}+\Phi_{1}\Phi_{1}),
\end{equation}
\begin{equation}\label{2.27}
v_{xy}-2|u|^{2}=0,
\end{equation}
\begin{equation}\label{2.28}
\Phi_{jxy}\Phi_{j}-\Phi_{jx}\Phi_{jy}+0.5|u|^{2}\Psi_{j}^{2}=0,
\end{equation}
\begin{equation}\label{2.29}
\Phi_{jxy}\Psi_{j}-\Psi_{jx}\Psi_{jy}+|u|^{2}\Psi_{j}^{2}=0,
\end{equation} 
where $j=1,2,3,4$.
\subsubsection{Ishimori  equation  with self-consistent sources}
The Ishimori  equation  with self-consistent sources was constructed in \cite{IshSCS}.
\subsection{Discrete equations}
\subsubsection{Toda Lattice Equation with Self-Consistent Sources }

The Toda lattice equation with $N$ self-consistent
sources (TLSCS)  is given by \cite{TLSCS}
\begin{equation}\label{2.30}
    v_t=v\bigg(p^{(-1)}+
    \sum_{j=1}^N\phi_{j-}^{(-1)}\phi_{j+}^{(-1)}\bigg)
    -v\bigg(p+\sum_{j=1}^N\phi_{j-}\phi_{j+}\bigg), 
    \end{equation}
    \begin{equation}\label{2.31}
		p_t=v\bigg(1+\sum_{j=1}^N\phi_{j-}^{(-1)}\phi_{j+}\bigg)-
    v^{(1)}\bigg(1+\sum_{j=1}^N\phi_{j-}\phi_{j+}^{(1)}\bigg),
    \end{equation}
    \begin{equation}\label{2.32}
		L\phi_{j+} = \lambda_j\phi_{j+}, \end{equation}
    \begin{equation}\label{2.33}
		L^* \phi_{j-} = \lambda_j\phi_{j-},\quad j=1,\ldots, N.
\end{equation}

The corresponding LR reads as
\begin{equation}\label{2.34}
      L\psi = v^{(1)}\psi^{(1)} + p\psi +\psi^{(-1)} =
    \lambda \psi \end{equation}
		\begin{equation}\label{2.35}
    -\psi_t = v^{(1)}\psi^{(1)} + \sum_{j=1}^N
    \frac{1}{\lambda-\lambda_j}v^{(1)}\phi_{j-}\left(\phi_{j+}^{(1)}\psi
      - \phi_{j+} \psi^{(1)}\right)\end{equation}
      \begin{equation}\label{2.36}
    L \phi_{j+} 
    = \lambda_j\phi_{j+}, 
    \end{equation}
    \begin{equation}\label{2.37}
    L^* \phi_{j-} 
    = \lambda_j\phi_{j-},\quad j=1,\ldots, N.
    \end{equation}

After the transformation
\begin{equation}
  \label{2.38}
  v:=\exp(x^{(-1)}-x),\quad
  p:= x_t - \sum_{j=1}^N\phi_{j+}\phi_{j-},
\end{equation}
we have \begin{equation}\label{2.39}
    x_{tt}=\exp\left(x^{(-1)}-x\right)
  \bigg(1+\sum_{j=1}^N\phi_{j+}\phi_{j-}^{(-1)}\bigg)
  -\exp\left(x-x^{(1)}\right)
  \bigg(1+\sum_{j=1}^N\phi_{j+}^{(1)}\phi_{j-}\bigg)
  +\sum_{j=1}^N\left(\phi_{j+}\phi_{j-}\right)_t.
\end{equation}

\subsubsection{Discrete KP equation with self-consistent sources}
The discrete KP equation with self-consistent sources were studied in \cite{1310.4636}
 
\subsection{Dispersionless equations}
\subsubsection{Dispersionless KP equation with
self-consistent sources}
The dispersionless KP equation with self-consistent sources (dKPSCS) has the form \cite{0510021}
\begin{subequations}
\label{2.40}
\begin{equation}
\label{2.40a}
     (u_{t}-3uu_{x}+\sum_{i=1}^Nv_{ix})_x=\frac{3}{4}u_{yy},
\end{equation}
\begin{equation}
\label{2.40b}
     p_{iy}=(p_i^2+2u)_x,
\end{equation}
\begin{equation}
\label{2.40c}
    v_{iy}=2(v_ip_i)_x,\ \ i=1,..., N.
\end{equation}
\end{subequations}
It is  the compatibility of the
following  equations 
\begin{subequations}
\label{2.41}
\begin{equation}
\label{2.41a}
     p_y=(p^2+2u)_x=2pp_x+2u_{x},
\end{equation}
\begin{equation}
\label{2.41b}
p_t=(p^3+3up+3w+\sum_{i=1}^N\frac{v_i}{p-p_i})_x=3p^2p_x+3(up)_x+3w_{x}+\sum_{i=1}^N\frac{v_{ix}}{p-p_i}-\sum_{i=1}^N\frac{v_i(p_x-p_{ix})}{(p-p_i)^2},
\end{equation}
\end{subequations}
where $w_{x}=\frac{1}{2}u_{y}$.

\subsubsection{Dispersionless mKP equation with
self-consistent sources}
The dmKP equation with self-consistent sources (dmKPESCS) can be written as \cite{0606060}
\begin{subequations}
\label{2.42}
\begin{equation}\label{2.42a}
     2v_t-\frac{3}{2}D_x^{-1}(v_{yy})-3v_xD_x^{-1}(v_y)+3v^2v_x-2\sum_{i=1}^{N}(\frac{a_i}{p_i})_x=0,
\end{equation}
\begin{equation}
\label{2.42b}
     a_{iy}=2[a_i(p_i+v)]_x,
\end{equation}
\begin{equation}
\label{2.42c}
    p_{iy}=(p_i^2+2vp_i)_x,\ \ i=1,..., N.
\end{equation}
\end{subequations}

\section{Integrable Heisenberg ferromagnetic equation  with self-consistent sources}
One of the important soliton equations is the Heisenberg ferromagnet equation (HFE) which can be written as \cite{lakshmanan77}-\cite{takhtajan}
\begin{equation}
\label{3.43}
iS_{t}+0.5[S,S_{xx}]=0,
\end{equation}
where 
\begin{equation}\label{3.44a}
S=
\left(\begin{array}{cc}
S_{3}&S^{-}\\
S^{+}&-S_{3}
\end{array}\right), \quad S^{+}S^{-}+S_{3}^{2}=1.
\end{equation}
It is integrable. In literature, some integrable and not integrable generalizations of the HFE (\ref{3.43}) were studied (see \cite{myrzakulov-391}-\cite{myrzakulov-1397} and references therein). Below we present the one of the integrable generalizations of the HFE (3.43) namely the integrable Heisenberg ferromagnetic equation  with self-consistent sources (HFESCS).
\subsection{The equation}Let us consider the Myrzakulov-XCIX (M-XCIX) equation \cite{1301.1649}-\cite{1404.2270}. It can be written in the different forms. One of its  form  reads as 
\begin{equation}
\label{3.44}
iS^{+}_{t}+S^{+}S_{3xx}-S^{+}_{xx}S_{3}+\frac{2}{a}\left(S^{+}\psi_{1}\psi_{2}-S_{3}\psi_{2}^{2}\right)=0,
\end{equation}
\begin{equation}
\label{3.45}
iS^{-}_{t}-(S^{-}S_{3xx}-S^{-}_{xx}S_{3})-\frac{2}{a}\left(S^{-}\psi_{1}\psi_{2}+S_{3}\psi_{1}^{2}\right)=0,
\end{equation}
\begin{equation}
\label{3.46}
iS_{3t}+0.5(S^{-}S_{xx}^{+}-S^{-}_{xx}S^{+})+\frac{1}{a}\left(S^{-}\psi_{2}^{2}+S^{+}\psi_{1}^{2}\right)=0,
\end{equation}
\begin{equation}
\label{3.47}
\psi_{1x}-ia(S_{3}\psi_{1}+S^{-}\psi_{2})=0,
\end{equation}
\begin{equation}
\label{3.48}
\psi_{2x}-ia(S^{+}\psi_{1}-S_{3}\psi_{2})=0,
\end{equation}
where $a$ is a complex constant,
$\psi=(\psi_{1},\psi_{2})^{T}$ and 
$S^{+}S^{-}+S^{2}_{3}=1$. 
Note that this  system  we   can  rewrite  equivalently as
\begin{equation}
\label{3.49}
iS^{+}_{t}+\left(S^{+}S_{3x}-S^{+}_{x}S_{3}-ia^{-2}\psi_{2}^{2}\right)_{x}=0,
\end{equation}
\begin{equation}
\label{3.50}
iS^{-}_{t}-\left(S^{-}S_{3x}-S^{-}_{x}S_{3}+ia^{-2}\psi_{1}^{2}\right)_{x}=0,
\end{equation}
\begin{equation}
\label{3.51}
iS_{3t}+0.5\left(S^{-}S_{x}^{+}-S^{-}_{x}S^{+}-2ia^{-2}\psi_{1}\psi_{2}\right)_{x}=0,
\end{equation}
\begin{equation}
\label{3.52}
\psi_{1x}-ia(S_{3}\psi_{1}+S^{-}\psi_{2})=0,
\end{equation}
\begin{equation}
\label{3.53}
\psi_{2x}-ia(S^{+}\psi_{1}-S_{3}\psi_{2})=0.
\end{equation}
The above system we can interpretate as the HFESCS.
\subsection{The Lax represenation }

The LR of  the M-XCIX equation  looks like 
\begin{equation}
\label{3.54}
\Phi_x=U \Phi,
\end{equation}
\begin{equation}
\label{3.55}
\Phi_{t}=V\Phi,
\end{equation}
where
\begin{equation}\label{3.56}
U=-i\lambda S, \quad 
S=
\left(\begin{array}{cc}
S_{3}&S^{-}\\
S^{+}&-S_{3}
\end{array}\right),\quad 
V=-2i\lambda^2S+\lambda SS_{x}+\left(\frac{i}{\lambda+a}-\frac{i}{a}\right)\left(\begin{array}{cc}
\psi_{1}\psi_{2}&-\psi_{1}^{2}\\
\psi_{2}^{2}&-\psi_{1}\psi_{2}
\end{array}\right).
\end{equation}
\subsection{Gauge equivalent counterpart}
The gauge equivalent counterpart of the M-XCIX equation  is given by
\begin{equation}\label{3.57}
iq_{t}+q_{xx}+2q^2r+2i\varphi_{1}^2=0,\end{equation}
\begin{equation}\label{3.58}
ir_{t}-r_{xx}-2qr^2+2i\varphi_{2}^{2}=0,
\end{equation}
\begin{equation}
\label{3.59}
\varphi_{1x}-ia\varphi_{1}-q\varphi_{2}=0,
\end{equation}
\begin{equation}
\label{3.60}
\varphi_{2x}+r\varphi_{1}+ia\varphi_{2}=0,
\end{equation}
where $\varphi=(\varphi_1,\varphi_2)^T$. It is the   AKNSSCS (2.19)-(2.21) for the one source case \cite{0412071}. The LR of  the AKNSES  reads as \cite{0412071}
\begin{equation}\label{3.61}
\psi_x=U\psi,
\end{equation}
\begin{equation}\label{3.62}
\psi_{t}=V\psi,
\end{equation}
where
\begin{equation}\label{3.63}
U=
\left(\begin{array}{cc}
-\lambda&q\\
-r&\lambda
\end{array}\right),\quad
V=i\left(\begin{array}{cc}
-2\lambda^2+qr&2\lambda q+q_x\\
-2\lambda r+r_x&2\lambda^2-qr
\end{array}\right)+\frac{i}{\lambda+a}
\left(\begin{array}{cc}
-\varphi_{1}\varphi_{2}&\varphi_{1}^2\\
-\varphi_{2}^2&\varphi_{1}\varphi_{2}
\end{array}\right).
\end{equation}

 \section{Some  generalizations of the HFESCS}
 If we have $N$ sources, then the HFESCS (3.44)-(3.48) takes the form
\begin{equation}
\label{4.64}
iS^{+}_{t}+S^{+}S_{3xx}-S^{+}_{xx}S_{3}+2\sum_{j=1}^{N}a^{-1}_{j}\left(S^{+}\psi_{1j}\psi_{2j}-S_{3}\psi_{2j}^{2}\right)=0,
\end{equation}
\begin{equation}
\label{4.65}
iS^{-}_{t}-(S^{-}S_{3xx}-S^{-}_{xx}S_{3})-2\sum_{j=1}^{N}a^{-1}_{j}\left(S^{-}\psi_{1j}\psi_{2j}+S_{3}\psi_{1j}^{2}\right)=0,
\end{equation}
\begin{equation}
\label{4.66}
iS_{3t}+0.5(S^{-}S_{xx}^{+}-S^{-}_{xx}S^{+})+\sum_{j=1}^{N}a^{-1}_{j}\left(S^{-}\psi_{2j}^{2}+S^{+}\psi_{1j}^{2}\right)=0,
\end{equation}
\begin{equation}
\label{4.67}
\psi_{1jx}-ia_{j}(S_{3}\psi_{1j}+S^{-}\psi_{2j})=0,
\end{equation}
\begin{equation}
\label{4.68}
\psi_{2jx}-ia_{j}(S^{+}\psi_{1j}-S_{3}\psi_{2j})=0.
\end{equation}
We can rewrite this equation also as
\begin{equation}
\label{4.69}
iS^{+}_{t}+\left(S^{+}S_{3x}-S^{+}_{x}S_{3}-i\sum_{j=1}^{N}a^{-2}_{j}\psi_{2j}^{2}\right)_{x}=0,
\end{equation}
\begin{equation}
\label{4.70}
iS^{-}_{t}-\left(S^{-}S_{3x}-S^{-}_{x}S_{3}+i\sum_{j=1}^{N}a^{-2}_{j}\psi_{1j}^{2}\right)_{x}=0,
\end{equation}
\begin{equation}
\label{4.71}
iS_{3t}+0.5\left(S^{-}S_{x}^{+}-S^{-}_{x}S^{+}-2i\sum_{j=1}^{N}a^{-2}_{j}\psi_{1j}\psi_{2j}\right)_{x}=0,
\end{equation}
\begin{equation}
\label{4.72}
\psi_{1jx}-ia_{j}(S_{3}\psi_{1j}+S^{-}\psi_{2j})=0,
\end{equation}
\begin{equation}
\label{4.73}
\psi_{2jx}-ia_{j}(S^{+}\psi_{1j}-S_{3}\psi_{2j})=0.
\end{equation}
The LR of this generalized HFESCS has the form
\begin{equation}
\label{4.74}
\Phi_x=-U \Phi,
\end{equation}
\begin{equation}
\label{4.75}
\Phi_{t}=V\Phi,
\end{equation}
where
\begin{equation}\label{4.76}
U=-i\lambda S, \quad 
S=
\left(\begin{array}{cc}
S_{3}&S^{-}\\
S^{+}&-S_{3}
\end{array}\right),\quad 
V=-2i\lambda^2S+\lambda SS_{x}+\sum_{j=1}^{N}\left(\frac{i}{\lambda+a_{j}}-\frac{i}{a_{j}}\right)\left(\begin{array}{cc}
\psi_{1j}\psi_{2j}&-\psi_{1j}^{2}\\
\psi_{2j}^{2}&-\psi_{1j}\psi_{2j}
\end{array}\right).
\end{equation}

 \section{Relation between soliton equations with self-consistent potentials and  soliton equations with self-consistent sources}
 As well-known, there are exist several soliton equations with self-consistent potentials. It is interesting to note that  these equations are related with SESCS. Here we demonstrate these relations in  two examples.
 
\subsection{The M-XCIX equation} 
The standard form of the M-XCIX equation reads as \cite{1301.1649}-\cite{1404.2270}
\begin{equation}
\label{5.77}
iS_{t}+0.5[S,S_{xx}]+a^{-1}[S,W]=0,
\end{equation}\begin{equation}
\label{5.78}
iW_{x}+a[S,W]=0.
\end{equation}
Let us solve the equation (5.78). Its solution (or one of the solutions) we can write as
\begin{equation}
W=\left(\begin{array}{cc}\label{5.79}
\psi_{1}\psi_{2}&-\psi_{1}^{2}\\
\psi_{2}^{2}&-\psi_{1}\psi_{2}
\end{array}\right),
\end{equation}
where $\psi_{j}$ are the solutions of the system (3.47)-(3.48). In terms of $S^{+}, S^{-}, S_{3}$ and $\psi_{j}$, the M-XCIX equation (5.77)-(5.78) takes the form  (3.44)-(3.48). 

\subsection{The NLS-Maxwell-Bloch equation} 
 
 Now we  return to  the AKNSSCS (3.57)-(3.60). Let us  we introduce 3 new functions as
 \begin{equation}\label{5.80}
p=-\varphi_{1}^2,\quad k=\varphi_{2}^{2}, \quad \eta=-\varphi_{1}\varphi_{2}.
\end{equation}
Then in terms of these new functions the AKNSSCS (3.57)-(3.60) takes the form
\begin{equation}\label{5.81}
iq_{t}+q_{xx}+2q^2r-2ip=0,\end{equation}
\begin{equation}\label{5.82}
ir_{t}-r_{xx}-2qr^2-2ik=0,
\end{equation}
\begin{equation}\label{5.83}
p_{x}-2iap-2q\eta=0,
\end{equation}
\begin{equation}
\label{5.84}
k_{x}+2iak-2r\eta=0,
\end{equation}
\begin{equation}
\label{5.85}
\eta_{x}+rp+qk=0,
\end{equation}
 It is nothing but the nonlinear Schrodinger-Maxwell-Bloch equation (see e.g. \cite{1301.1649}-\cite{1404.2270} and references therein). Now let us consider the reductions  $k=\delta \bar{p}, \quad r=\delta\bar{q}, \quad \delta=\pm 1$. Then the nonlinear Schrodinger-Maxwell-Bloch equation  takes the form
\begin{equation}\label{5.81}
iq_{t}+q_{xx}+2\delta|q|^2q-2ip=0,\end{equation}
\begin{equation}\label{5.83}
p_{x}-2iap-2q\eta=0,
\end{equation}
\begin{equation}
\label{5.85}
\eta_{x}+\delta(\bar{q}p+q\bar{p})=0.
\end{equation}
 \section{Conclusion}

 In this paper, we have briefly review of soliton equations with self-consistent sources. We have also presented their Lax  representations where we can. Then the integrable extension of the HFE namely the HFE with self-consistent sources are presented. Its Lax representation and gauge equivalent counterpart are also presented. Finally the integrable HFESCS with $N$ - sources with its LR are given. Finally we consider the relation between soliton equations with self-consistent potentials and  soliton equations with self-consistent sources.

 \end{document}